\documentclass[%
 aip,
 rsi,
amsmath,amssymb,
nobibnotes, 
reprint,
longbibliography
]{revtex4-1}

\usepackage{graphicx}
\usepackage{dcolumn}
\usepackage{bm}

\usepackage[utf8]{inputenc}
\usepackage{mathptmx}
\usepackage{multirow}
\usepackage{braket}
\usepackage{tablefootnote}

\usepackage[english]{babel}

\usepackage{xcolor}
\colorlet{RED}{red}
\colorlet{BLUE}{blue}
\usepackage{dcolumn}
\usepackage{bm}
\usepackage[version=4]{mhchem} 
\usepackage{acronym}
\usepackage{adjustbox}
\usepackage{float}
\usepackage{tikz}
\usepackage{microtype} 
\usepackage{algorithm}
\usepackage{enumitem}
\usepackage{appendix}
\usepackage{amsmath}
\usepackage{algpseudocode}
\usepackage[caption=false]{subfig}

\usetikzlibrary{calc,shapes.geometric,decorations.pathmorphing,patterns}

\definecolor{background-color}{gray}{0.98}
\usepackage[margin=2.3cm,bmargin=1cm,footnotesep=1cm]{geometry}

\usepackage{hyperref}
\hypersetup{
  colorlinks=true,
  linkcolor=blue,
  citecolor=blue,
  urlcolor=blue,
}

\begin{document}

\title{
Coupled Cluster Downfolding Theory in 
Simulations of Chemical Systems on Quantum Hardware
}

\author{Nicholas P. Bauman}
\email{nicholas.bauman@pnnl.gov}
\affiliation{%
  Physical Sciences Division,
  Pacific Northwest National Laboratory, Richland, Washington, 99354, USA
}

\author{Muqing Zheng}
\altaffiliation{Both N.P.B. and M.Z. contributed equally to this work}
\affiliation{%
Advanced Computing, Mathematics, and Data Division,
  Pacific Northwest National Laboratory, Richland, Washington, 99354, USA
}

\author{Chenxu Liu}
\affiliation{%
Advanced Computing, Mathematics, and Data Division,
  Pacific Northwest National Laboratory, Richland, Washington, 99354, USA
}

\author{Nathan M. Myers}
\affiliation{%
Advanced Computing, Mathematics, and Data Division,
  Pacific Northwest National Laboratory, Richland, Washington, 99354, USA
}

\author{Ajay Panyala}
\affiliation{%
Advanced Computing, Mathematics, and Data Division,
  Pacific Northwest National Laboratory, Richland, Washington, 99354, USA
}

\author{Bo Peng}
\affiliation{%
  Physical Sciences Division,
  Pacific Northwest National Laboratory, Richland, Washington, 99354, USA
}

\author{Ang Li}
\affiliation{%
  Advanced Computing, Mathematics, and Data Division,
  Pacific Northwest National Laboratory, Richland, Washington, 99354, USA
}
\affiliation{
Department of Electrical and Computer Engineering, University of Washington, Seattle, WA 98195, USA
}

\author{Karol Kowalski}
\email{karol.kowalski@pnnl.gov}
\affiliation{%
  Physical Sciences Division,
  Pacific Northwest National Laboratory, Richland, Washington, 99354, USA
 }
 \affiliation{%
  Department of Physics, University of Washington, Seattle, Washington 98195, USA
 }

\date{June 2025}

\maketitle

\noindent
\textbf{Abstract} 
The practical application of quantum technologies to chemical problems faces significant challenges, particularly in the treatment of realistic basis sets and the accurate inclusion of electron correlation effects. A direct approach to these problems is currently infeasible due to limitations in the number of logical qubits, their fidelity, and the shallow circuit depths supported by existing hardware; all of which hinder simulations at the required level of accuracy.
A promising alternative is hybrid quantum-classical computing, where classical resources are used to construct effective Hamiltonians characterized by dimensions that conform to the constraints of current quantum devices. In this paper, we demonstrate the performance of a hybrid approach: coupled-cluster downfolded Hamiltonians are first evaluated in reduced-dimensionality active spaces, and the corresponding ground-state energies are subsequently computed using quantum algorithms.
Our comprehensive analysis explores the achievable accuracy in recovering correlation energies when hundreds of orbitals are downfolded into a problem size tractable by today’s quantum hardware. We argue that such flexible hybrid algorithms, where problem size can be tailored to available quantum resources, can serve as a bridge between noisy intermediate-scale quantum (NISQ) devices and future fault-tolerant quantum computers, marking a step toward the early realization of quantum advantage in chemistry.
\\

\section{Introduction}

The accurate characterization  of the complex correlated behavior of electrons 
in molecules 
emerges as one of the most pressing challenges in the many-body theory of interacting systems.
Several classes  of methodologies, 
based on various representations of quantum mechanics including wave function \cite{coester58_421,cizek66_4256,raghavachari89_479,Bartlett2007}, electron density \cite{lee1988development,perdew1996generalized,jones2015density}, density matrix \cite{knizia2012density,wouters2016practical}, and Green's function \cite{hedin1965new,aryasetiawan1998gw,van2015gw,rehr2010parameter} 
based methods have been introduced, tested, and validated to address this problems.  Significant progress has been achieved in density functional theory \cite{teale2022dft}, density matrix renormalization group (DMRG) \cite{menczer2024parallel}, coupled cluster (CC) formalisms \cite{haugland2020coupled,yuwono2024relativistic,windom2024new,gururangan2025extension}, as well as variational formulations \cite{tubman2016deterministic,schriber2016communication,abraham2020selected,abraham2022coupled} and self-energy-based approaches \cite{van2006quasiparticle,deslippe2012berkeleygw,reeves2025performance}.  
Applications of these methods go well beyond quantum chemistry - analogous formulations are currently used in the context of materials sciences and nuclear physics (see, for example, Refs.\citenum{booth2013towards,zhang2019coupled,mcclain2017gaussian,ye2024periodic,hagen2016structure}). 

High-accuracy results generated by hierarchical classes of approximations -i.e., sequences of approximations that can reach the exact diagonalization limit - are usually associated with significant computational overhead and memory requirements as the system size increases, due to the large number of parameters needed to define accurate wave function expansions.
%
Quantum computing provides an alternative computational model to bypass the typical challenges of conventional computing associated with speed, memory, and energy limitations needed to achieve the so-called chemical accuracy. 
Algorithms such as Quantum Phase Estimation (QPE) \cite{nielsen2002quantum,kitaev1995quantum,Kitaev_1997,Abrams1999QPE,childs2010relationship,reiher2017elucidating} have theoretically the potential to overcome the exponential computational barriers preventing classical computing from reaching the exact diagonalization limit in simulations for realistic systems. 
However, for this to happen, the quantum hardware must mature to the point where deep quantum circuits and quantum error correction can be effectively implemented at scale.  Given the existing limitations of the quantum hardware, the hardware realization of QPE-type algorithms is scarce and still limited to the small-size, toy systems
\cite{aspuru2005simulated,wecker2014gate,wiebe2016efficient,santagati2018witnessing,mcardle2020quantum,klymko2022real}.  
For this reason, simpler algorithms 
are currently being used. An excellent example is provided by the class of hybrid Variational Quantum Eigensolvers (VQE)\cite{peruzzo2014variational,mcclean2016theory,romero2018strategies,PhysRevA.95.020501,Kandala2017,kandala2018extending,PhysRevX.8.011021,huggins2020non,cao2019quantum,weaving2025contextual}. Although hardware VQE executions for small chemical systems have been reported, their applications are also limited by circuit complexity and the need for explicit inclusion of a large number of wave function parameters, which makes VQE approaches not scalable.  Although hybrid  VQE applications can be performed for larger systems than possible with the QPE approach, their applications are limited to small-size active spaces, which in the best-case scenario can only encapsulate part of the static correlation effects. However, without an efficient way of incorporating dynamical correlation effects, usually associated with virtual orbitals falling outside of the active spaces, the accuracy of both QPE and VQE formalisms falls short of recovering the total correlation energies comprising both static and dynamical components needed to obtain an adequate level of accuracy in simulations for chemical processes. 

A practical approach for tackling these challenges while simultaneously providing a strategy for utilizing the resources of current quantum computers is linked to recent advancements in various dimensionality-reduction techniques of quantum problems, hereafter referred to as downfolding formalism \cite{bauman2019downfolding,downfolding2020t,huang2023leveraging}.
These formulations lay the foundation for efficient hybrid quantum/classical execution protocols. For example, the recently introduced coupled cluster downfolding formalisms allow the construction of effective Hamiltonians that encapsulate dynamical correlation effects within the active spaces whose dimensions can be tuned to the available quantum resources. The CC downfolding introduced in Refs.\citenum{kowalski2018properties,bauman2019downfolding,downfolding2020t,kowalski2021dimensionality,huang2023leveraging,feldmann2024complete} or other embedding techniques based on the CC downfolding techniques~\cite{peng2024integrating,shee2024static} utilize the simplicity of single-reference unitary  CC methods and provide a platform for hierarchical inclusion of collective many-body effects. 
 The CC downfolding method, in conjunction with quantum solvers based on the exponential form of the wave function ansatz (i.e. unitary CC formulations~\cite{hoffmann1988unitary,unitary1,lee2018generalized}), in contrast to quantum algorithms targeting some form of truncated configuration interaction formalism, provides size-consistency or size-extensivity of calculated energies (i.e., additive separability of energy in the noninteracting subsystem limit), a property that has been identified by John Pople in his Nobel Lecture \cite{pople1999nobel} as {\it a matter of great importance}. We argue that satisfying this requirement should be addressed at the early stages of applications of quantum computing in chemistry.
Furthermore, its multi-active space variant, referred to as the Quantum Flow approach (QFlow) \cite{kowalski2023quantum,kowalski2025resource}, facilitates the exploration of extensive subspaces within Hilbert space through the use of reduced quantum resources and constant-depth quantum circuits, giving rise to efficient utilization of distributed quantum resources. 
Some QFlow formulations require no prior knowledge about the structure of the sought-after state.

\begin{figure*}
    \includegraphics[width=\linewidth]{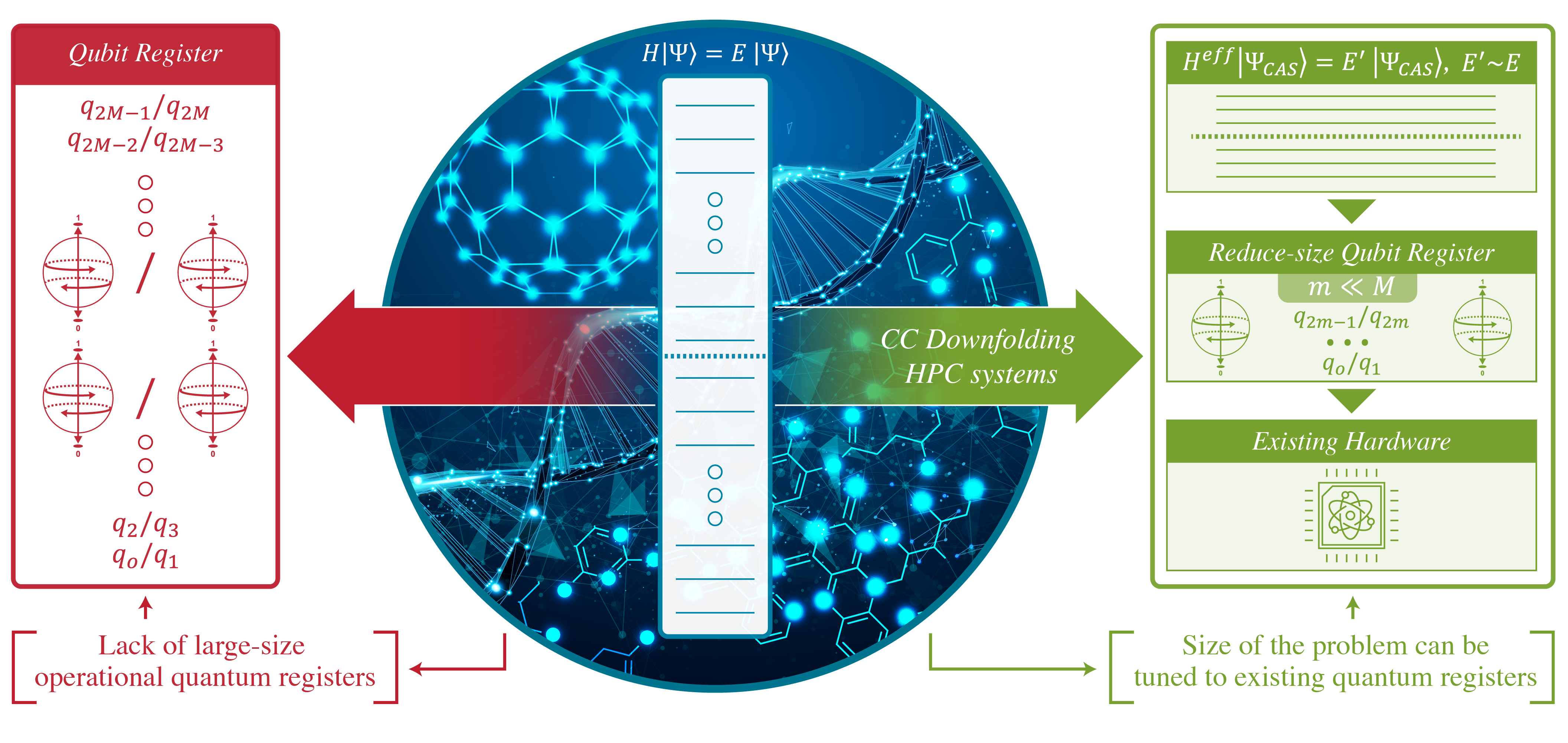}
    \caption{{\bf Coupled cluster downfolding formalism to reduce the dimensionality of the quantum problem.} 
%
Brute-force approaches for solving electronic Hamiltonians defined by large basis sets are not feasible with current quantum hardware. As an alternative, a downfolding procedure can be employed to construct effective (downfolded) Hamiltonians ($H^{\rm eff}$) within low-dimensional complete active spaces (CASs) that are suitable for existing quantum computers. This approach offers a flexible framework that can adapt to rapidly evolving quantum technologies by increasing the size of the active space. The effective Hamiltonians capture both in- and out-of-active-space electron correlation effects and enable accurate energy calculations using reduced quantum resources with current quantum algorithms such as VQE and GCIM to approximate their ground-state eigenvectors $|\Psi_{CAS}\rangle$. The construction of downfolded Hamiltonians requires classical computing resources, ranging from leadership-class supercomputers to cloud-computing systems.
}
    \label{fig1}
\end{figure*}

Here, we demonstrate the performance of a hybrid computational pipeline called Quantum Infrastructure for Reduced-Dimensionality Representations (QRDR). This pipeline integrates classical and quantum computing resources into a flexible infrastructure that adapts to rapidly evolving quantum hardware. It consists of three main components.
First, the pipeline employs highly scalable codes to compute downfolded Hamiltonians.
Second, it uses four quantum solvers: ADAPT-VQE, qubit-ADAPT-VQE, the generator-coordinate-inspired method (GCIM) \cite{zheng2023quantum, zheng2024unleashed}, and a VQE based on the generalized unitary coupled cluster ansatz (UCCGSD) \cite{lee2018generalized} to optimize downfolded Hamiltonians within active spaces on selected backends. Those backends include quantum hardware and our third component, the SV-Sim state-vector simulator \cite{li2021svsim} that is specialized for efficient circuit simulation on high-performance computing (HPC) systems.
This framework, schematically shown in Fig.~\ref{fig1}, is subsequently applied to three molecular systems where correlation effects cannot be accurately captured using the conventional bare Hamiltonian simulations in active space, a prevalent model in simulating chemistry on quantum computers. In particular, we demonstrate that QRDR outperforms the above-mentioned approach by incorporating correlation effects into the active space in an elegant and hierarchical manner. We also illustrate the accuracy amplification mechanism for energies associated with classical simulations used to construct downfolded Hamiltonians and their subsequent optimization on quantum hardware.

\section{Results}
\label{section:results}
\begin{figure}
    \includegraphics[width=\linewidth]{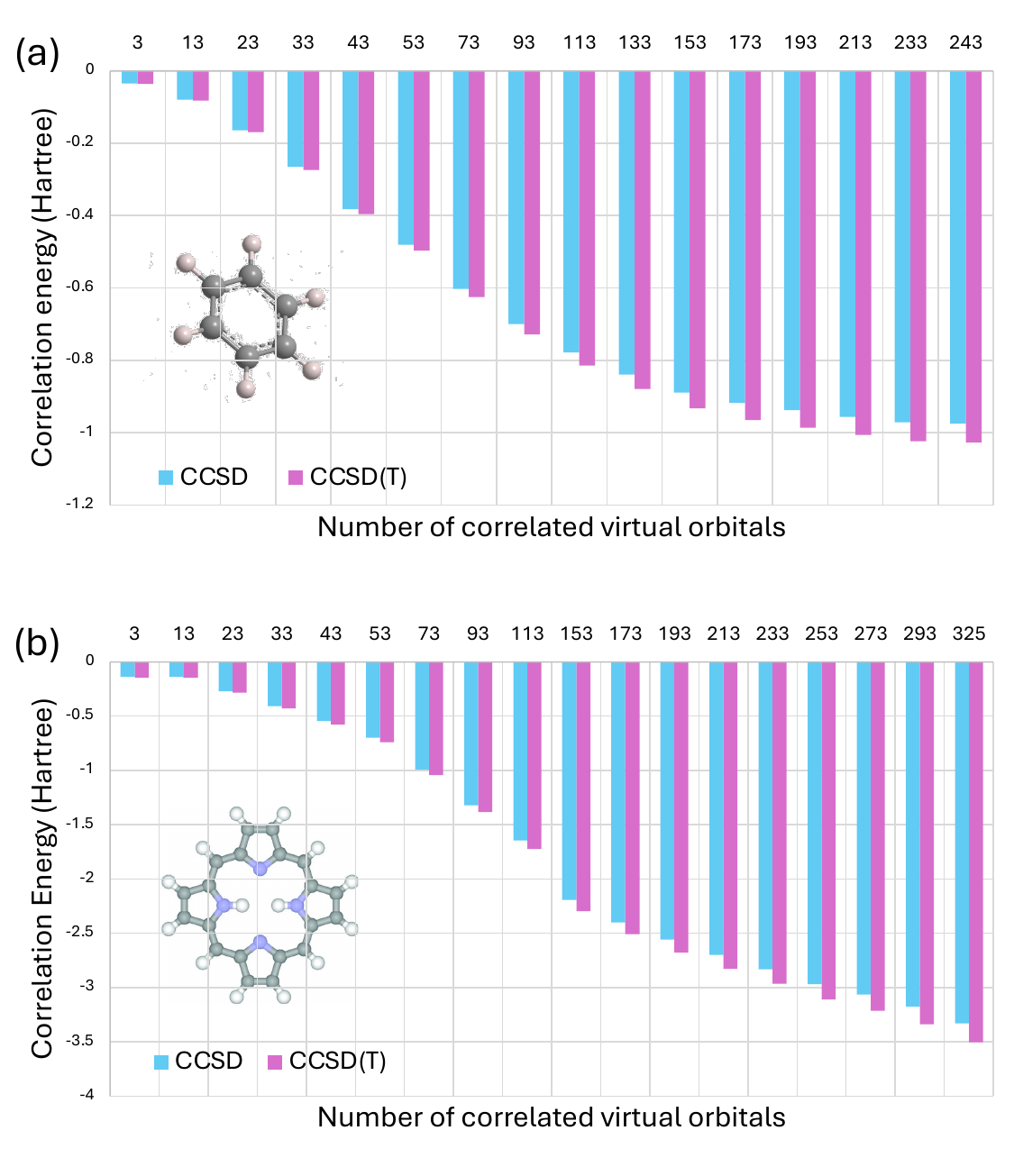}
    \caption{Structure of the correlation effects in the benzene molecule (inset (a)) and free-base porphyrin (inset (b)), using the cc-pVTZ and cc-pVDZ basis sets, respectively, at the CCSD and CCSD(T) levels of theory, as a function of the number of correlated virtual orbitals (all core orbitals were frozen during the calculations).}
    \label{benzene_fbp_correlation}
\end{figure}

We performed QRDR simulations for several benchmark systems, including those dominated by strong dynamical correlation effects, which in the brute-force quantum computing approach would require the use of an excessive number of qubits to capture these effects. 
As benchmark systems, we chose N$_2$ in cc-pVTZ \cite{dunning1989gaussian},  benzene molecule in the  cc-pVDZ \cite{dunning1989gaussian} and cc-pVTZ, and free-base porphyrin molecule (FBP) in the cc-pVDZ basis sets:\\
\begin{itemize}
\item{\bf N$_2$ system:} The balance of the correlation effects vary with the N-N distance from dominating dynamical  (equilibrium) to the significant static (stretched N-N bond) components. 
\item{\bf Benzene molecule:} for the C$_6$H$_6$ system at the equilibrium geometry the correlation effects are dominated by its dynamical component. As shown in Fig.~\ref{benzene_fbp_correlation}(a), the correlation energy grows almost linearly with the number of correlated virtual orbitals.
\item{\bf Free-base porphyrin:} The C$_{20}$H$_{14}$N$_4$ system, in analogy to the benzene case,  is characterized by the presence of strong dynamical correlation effects amounting to nearly 3.5 Hartree of correlation energy. As for the benzene molecule, the correlation energy depends almost linearly on the number of correlated virtual orbitals (see Fig.~\ref{benzene_fbp_correlation}(b)). 
\end{itemize}
In our simulations, we employed Cartesian geometries for benzene and FBP used in Refs.~\onlinecite{eriksen2020ground} and  \onlinecite{kowalski2010active}, respectively. For N$_2$ the equilibrium geometry R$_e$ is assumed to be 2.068 $a_0$. In N$_2$ simulations, we correlated all molecular orbitals, while in simulations for the benzene and FBP systems, we kept all core orbitals frozen.
Table \ref{tab:classical} presents ground-state energies obtained using several quantum chemical methods implemented on classical computers, including restricted Hartree--Fock (RHF), CCSD \cite{purvis82_1910}, CCSD(T) \cite{raghavachari89_479}, and CCSDTQ \cite{Kucharski1991,ccsdtq_nevin,piecuch1994state} approaches, to evaluate the accuracy of the results from quantum hardware. For the benzene molecule in the cc-pVDZ basis set, a comparison of the CCSDTQ energies with other formulations indicates that the CCSD(T) approach accurately captures the non-dynamical correlation energy, which is the dominant contribution to the total correlation energy.


The level of correlation energy recovery in typical active-space simulations based on the bare Hamiltonian is presented in Table \ref{tab:corr-energy}, where we compare the correlation energies obtained from CCSD(T) calculations in the ($6e,6o$) active space (hereafter referred to as CCSD(T)($6e,6o$)) with those from full CCSD(T) calculations using the frozen-core approximation.
For the cc-pVDZ C$_6$H$_6$ model, the CCSD(T)($6e,6o$) calculation recovers approximately $3 \%$ of the total CCSD(T) correlation energy. In contrast, for the larger cc-pVDZ C$_{20}$H$_{14}$N$_4$ system, the CCSD(T)($6e,6o$) method recovers less than $1 \%$ of the corresponding CCSD(T) correlation energy. These results clearly illustrate that bare-Hamiltonian active-space simulations are insufficient to achieve the desired level of accuracy in modeling chemical systems.
While one might argue that energy differences are typically more relevant in chemical simulations, the lack of a balanced and simultaneous treatment of static and dynamical correlation effects can still lead to significant inaccuracies.
For comparison, Table \ref{tab:corr-energy} also includes results from qubit-ADAPT-VQE hardware simulations ({\it vide infra}).

The effective Hamiltonians are constructed in active spaces defined by the RHF orbitals. For practical reasons, primarily due to the limitations of quantum hardware and the associated cost of accessing these systems, we opted to use relatively small active spaces, defined by six active electrons distributed over six active orbitals for all benchmark systems, i.e., ($6e,6o$) active spaces. The hybrid execution workflow comprises classical computations to generate effective Hamiltonians, followed by energy evaluation and optimization on  quantum simulators or quantum hardware. 

The energies evaluated on the noiseless simulators are presented in Table \ref{tab:simulator}, where they are compared with full configuration interaction (FCI) results obtained via exact diagonalization of downfolded Hamiltonians in 
($6e,6o$) active spaces. For larger systems dominated by dynamical correlation effects (i.e., benzene and FBP), we observe excellent agreement between the energies computed using all noiseless simulators and the corresponding FCI values. A similar level of agreement is seen for the N$_2$ system at its equilibrium
geometry (R$_{\rm N-N}$ = 1.0R$_e$). For larger bond stretches of the N–N distance, where static correlation effects become increasingly 
significant, we find that among all noiseless simulators, the ADAPT-GCIM(2,2) formulation of Ref. \citenum{zheng2024unleashed} and the UCCGSD approach provide the closest agreement with the FCI results.

%
Given the limited quantum resources, we focused exclusively on hardware simulations of the benzene and FBP systems. The ground-state energies obtained from hardware experiments for benzene (using the cc-pVTZ basis set) and FBP (using the cc-pVDZ basis set) are presented in Table \ref{tab:hardware}.
In all experiments, we employed downfolded Hamiltonians based on the double unitary coupled cluster ansatz (DUCC) consistent with the DUCC(3)-A(7) approximation described in Ref.\citenum{doublec2022}
(see Appendix \ref{appendix:a7} for further details).
For both systems, the energies measured on the Quantinuum H1-1 quantum computer show a good agreement with those obtained using the H1-1 Emulator for non-noise-amplified circuits. Specifically, the energy discrepancy for benzene is $3.4$ milliHartree, while for FBP it is $6.7$ milliHartree.

With the Zero Noise Extrapolation (ZNE) error mitigation method (outlined in Section \ref{sub:hardware}), the error-mitigated energy estimates from the Quantinuum H1-1 hardware closely match the {\em target energies}, which are defined using the CCSD(T) method. The difference between the H1-1 hardware result and the CCSD(T) benchmark is approximately $17.1$ milliHartree for benzene. Due to credit limitations, we did not have enough resources to perform ZNE for FBP. For unmitigated estimations, the differences to the CCSD(T) benchmark are $45.2$ milliHartree for Benzene and $66.5$ milliHartree for FBP. 
Notably, with or without error mitigation, almost all estimates from H1-1, $ibm\_marrakesh$, and $ibm\_kingston$ quantum computers in Table \ref{tab:hardware} significantly outperform the {\em source energies} 
derived from the CCSD method, which provides the external cluster amplitudes used to construct the downfolded Hamiltonians in the active spaces. We refer to this improvement as {\it accuracy amplification}, highlighting the effectiveness of the downfolding procedure in capturing out-of-active-space correlation effects.

\begin{table*}
   \centering
   \caption{Comparison of ground-state energies for the benchmark systems considered. In CC calculations for N$_2$, all orbitals were correlated, whereas for benzene and the FBP molecule, core electrons were frozen. All simulations were performed using the NWChem electronic structure software \cite{apra2020nwchem}.}
    \label{tab:classical}
    \begin{tabular}{ccccccc} \hline \hline \\
  System & Basis set  & Basis set size          &  RHF & CCSD  & CCSD(T)  &  CCSDTQ   \\ \hline \\[-0.1cm]
    N$_2$ ($1.0$R$_e$)      & cc-pVTZ & 60 &-108.9841  &  -109.3811 &  -109.3998  &  -109.4011    \\[0.2cm]
    N$_2$  ($1.5$R$_e$)     & cc-pVTZ & 60  & -108.5902   &  -109.1092  & -109.1616 &  -109.1677   \\[0.2cm]
    N$_2$   ($2.0$R$_e$)    & cc-pVTZ & 60 &-108.2651  & -108.9681    &   -109.1448     &  -109.0691   \\[0.2cm]
  C$_6$H$_6$  & cc-pVDZ & 114 &-230.7218   & -231.5453 & -231.5813 & -231.5842\footnote{CCSDTQ energy taken from Ref.\citenum{eriksen2020ground}.}   \\[0.2cm]
  C$_6$H$_6$  & cc-pVTZ & 264 &-230.7785   & -231.7537 & -231.8058 &    ---     \\[0.2cm] 
  C$_{20}$H$_{14}$N$_4$ & cc-pVDZ & 406 &   -983.3165   &  -986.6465         &      -986.8192  &    ---     \\
  \hline \hline
    \end{tabular}
\end{table*}

\begin{table*}
    \centering
    \caption{Comparison of correlation energies obtained using various many-body methods designed for classical computing—CCSD(T) in full space and in an active space [CCSD(T)($6e,6o$)]—with qubit-ADAPT-VQE (Q-A-VQE) energies simulated on quantum hardware. The CCSD(T)($6e,6o$) results demonstrate the accuracy level achievable in active-space simulations of bare Hamiltonians. All correlation energies are given in Hartrees.}
    \label{tab:corr-energy}
    \begin{tabular}{cccccc} \hline \hline \\
  System & Basis set  & CC($6e,6o$)   & H1-1 Hardware  &  CC  \\[0.2cm]
  \hline \\
  C$_6$H$_6$  & cc-pVTZ &  -0.031 & -1.010   & -1.027 \\[0.1cm]  
        &    & CCSD(T)($6e,6o$) & Q-A-VQE &   CCSD(T) \\[0.2cm]
\hline \\ 
        
  C$_{20}$H$_{14}$N$_4$  & cc-pVDZ & -0.033  &  -3.426  & -3.503 \\[0.1cm]  
        &    & CCSD(T)($6e,6o$) & Q-A-VQE &  CCSD(T) \\[0.2cm]
  \hline \hline
    \end{tabular}
\end{table*}

\begin{table*}
   \centering
   \caption{Ground-state energies obtained using noiseless simulators for the downfolded Hamiltonians represented in ($6e,6o$) active spaces. In all cases, the approximate form of the effective Hamiltonian given by Eq.~\eqref{a7exp} was used. The following solvers were employed: ADAPT-VQE (A-VQE),  qubit-ADAPT-VQE (Q-A-VQE),  ADAPT-GCIM (A-GCIM), ADAPT-GCIM(2,2) (A-GCIM(2,2)), and generalized UCCSD (UCCGSD). These results are compared with exact (FCI) values obtained from the exact diagonalization of the effective Hamiltonian within the corresponding ($6e,6o$) active spaces.}
    \label{tab:simulator}
    \begin{tabular}{cccccc} \hline \hline \\
 FCI & A-VQE & A-GCIM & A-GCIM(2,2) & Q-A-VQE & UCCGSD \\ \hline \\[-0.1cm]
\multicolumn{6}{c}{N$_2$, R$_{N-N}$=1.0R$_e$, cc-pVTZ} \\[0.2cm]
-109.3908 & -109.3908 & -109.3908 & -109.3908 & -109.3908 & -109.3908 \\
 \hline \\[-0.2cm]
 
\multicolumn{6}{c}{N$_2$, R$_{N-N}$=1.5R$_e$, cc-pVTZ}  \\[0.2cm]
-109.1303 & -109.1296 & -109.1303 & -109.1303 & -109.1297 & -109.1300 \\
 \hline \\[-0.2cm]

\multicolumn{6}{c}{N$_2$, R$_{N-N}$=2.0R$_e$, cc-pVTZ} \\[0.2cm]
-108.9842 & -108.9834 & -108.9823 & -108.9840 & -108.9834 & -108.9839 \\
\hline \\[-0.2cm]
 
\multicolumn{6}{c}{C$_6$H$_6$, cc-pVDZ} \\[0.2cm]
-231.5711 & -231.5711 & -231.5711 & -231.5711 & -231.5711 &  -231.5711 \\
\hline \\[-0.2cm]

\multicolumn{6}{c}{C$_6$H$_6$, cc-pVTZ} \\[0.2cm]
-231.7878 & -231.7878 & -231.7878 & -231.7878 & -231.7878 &  -231.7878 \\ \hline \\[-0.2cm]
\multicolumn{6}{c}{C$_{20}$H$_{14}$N$_4$, cc-pVDZ} \\[0.2cm] 
-986.7732 & -986.7731 & -986.7731 & -986.7731 & -986.7731 &  -986.7732 \\[0.1cm]
  \hline \hline
    \end{tabular}
\end{table*}

\begin{table*}
   \centering
   \caption{
Hardware simulations using Quantinuum and IBM quantum computers. One-shot DUCC(3) Qubit-ADAPT-VQE simulations performed on noise emulators and quantum hardware are presented.   
   \footnote{Experiment details are explained in Section~\ref{sub:hardware}. The uncertainty for each estimation is the standard error (SE).}.}
    \label{tab:hardware}
    \begin{tabular}{cccccc} \hline \hline \\
 \multicolumn{3}{c}{\hspace*{1.5cm} Benzene (cc-pVTZ)\hspace*{1.5cm}} &  \multicolumn{3}{c}{\hspace*{1.5cm} FBP (cc-pVDZ) \hspace*{1.5cm}} \\
\multicolumn{3}{c}{\hrulefill} &   \multicolumn{3}{c}{\hrulefill} \\
Quantinuum  & Quantinum        & ---            & Quantinuum  &  Quantinum & --- \\
H1-1 Emulator & H1-1 Hardware & ---          & H1-1 Emulator & H1-1 Hardware & --- \\
%
$-231.7572 \pm 0.0064$  & $-231.7606 \pm 0.0088$ &    &$-986.7495 \pm 0.0037$  &$-986.7428 \pm 0.0039$  &  \\[-0.1cm]
\multicolumn{3}{c}{\hrulefill} &   \multicolumn{3}{c}{\hrulefill} \\
Quantinuum  & Quantinum        & IBM            & Quantinuum  &  Quantinum & IBM\\
H1-1 Emulator+ZNE & H1-1 Hardware+ZNE & $marrakesh$ + QESEM          & H1-1 Emulator+ZNE & H1-1 Hardware+ZNE & $kingston$ + QESEM\\
%
$-231.7652 \pm 0.0054$ & $-231.7887 \pm 0.0072$ & $-231.7810 \pm 0.0259$          &$-986.7535 \pm 0.0030$ & --- & $-986.7369 \pm 0.0171$\\[-0.1cm]
\multicolumn{3}{c}{\hrulefill}                  &   \multicolumn{3}{c}{\hrulefill} \\
 \multicolumn{3}{c}{Classical source for external amplitudes: CCSD} &  \multicolumn{3}{c}{Classical source for external amplitudes: CCSD} \\
  \multicolumn{3}{c}{$-231.7537$} &  \multicolumn{3}{c}{$-986.6465$} \\[-0.2cm]
 \multicolumn{3}{c}{\hrulefill} &   \multicolumn{3}{c}{\hrulefill} \\
  \multicolumn{3}{c}{Classical target: CCSD(T)} &  \multicolumn{3}{c}{Classical target: CCSD(T)} \\
  \multicolumn{3}{c}{$-231.8058$} &  \multicolumn{3}{c}{$-986.8192$} \\
  \hline \hline
\end{tabular}
\end{table*}





\section{Methods and Computational Details}
\subsection{CC downfolding and effective Hamiltonians}


Coupled-cluster downfolding techniques have emerged as powerful tools for reducing the dimensionality of quantum many-body problems. Central to these approaches is the exponential CC ansatz, which facilitates the construction of effective (downfolded) Hamiltonians within a reduced-dimensional active space. These effective Hamiltonians encode the influence of out-of-active-space correlation effects (i.e., dynamical correlation) into their many-body structure. This formulation ensures that the lowest eigenvalue of the effective Hamiltonian corresponds to the exact or approximate CC energy.

Two primary variants of CC downfolding exist, based on the standard single-reference CC \cite{kowalski2018properties} and the more general unitary CC ansatz \cite{bauman2019downfolding}. The latter leads to Hermitian CC downfolding, which is particularly relevant for quantum computing applications. In contrast, non-Hermitian CC downfolding provides a mathematically elegant framework for connecting different active-space problems, an idea formalized in the Equivalence Theorem (see Ref. \citenum{kowalski2021dimensionality}). This connection underpins the concept of Quantum Flow \cite{kowalski2023quantum}, which is also explored within the Hermitian formalism.



Although the use of effective Hamiltonians has a long history in quantum chemistry and physics \cite{bloch1958theorie,des1960extension,lowdin1963studies,schrieffer1966relation,soliverez1969effective,jorgensen1975effective,glazek1993renormalization,bravyi2002fermionic,durand1983direct,kaldor1991fock,jezmonk,jeziorski1989valence,bernholdt1999critical,pal1988molecular,meissner1995effective,meissner1998fock,li2003general,mrcclyakh}, it has only recently been recognized that the effective Hamiltonian formalism arises naturally from the single-reference ansatz \cite{kowalski2018properties,bauman2019downfolding,downfolding2020t,kowalski2021dimensionality,kowalski2023sub}.
For example, in the non-Hermitian case, the lowest eigenvalues of the effective Hamiltonian reproduce the standard coupled-cluster (CC) energies that are computed using the textbook CC energy formula to numerical precision.


The Hermitian formulations utilize the DUCC ansatz\cite{bauman2019downfolding,downfolding2020t}, where the exact ground-state wave function is represented as the product of two exponential ansatzes defined by anti-Hermitian cluster operators $\sigma_{\rm ext}$ 
(similar decomposition of the standard cluster operator is employed in active-space CC formulations see Refs.\citenum{oliphant1991multireference,oliphant1992implementation,pnl93}
)
and $\sigma_{\rm int}$
that are expressed in terms of 
parameter containing all spin-orbital indices active ($\sigma_{\rm int}$) and  at least one spin-orbital index  inactive ($\sigma_{\rm ext}$).
Using the DUCC ansatz, one can construct (once certain conditions are met) an effective Hamiltonian $H^{\rm eff}$ in the active space
\begin{equation}
    H^{\rm eff} = (P+Q_{\rm int})
    e^{-\sigma_{\rm ext}} H
    e^{\sigma_{\rm ext}}
    (P+Q_{\rm int})\;.
    \label{eqeff}
\end{equation}
which can reproduce the lowest eigenvalue being exact (or approximate) energy $E$ ($E'$) of the system once exact (approximate) form of $\sigma_{\rm ext}$ is known. 
The $P+Q_{\rm int}$ is a projection operator onto active space
($P=|\Phi\rangle\langle\Phi|$  is a projection operator onto the reference function $|\Phi\rangle$ and $Q_{\rm int}$ represents a projection operator onto all excited configurations with respect to $|\Phi\rangle$ in active space).
Using commutator expansion for Eq. \eqref{eqeff} one can show that the  $H^{\rm eff}$ operator is expressed in terms of connected diagrams only. 

In general, due to the non-commutativity of the many-body components that define $\sigma_{\rm ext}$, analyzing the structure of the Hermitian effective Hamiltonian is more challenging than in the non-Hermitian case. Practical implementations require several approximations, including the treatment of the similarity transformation (see Eq.~(\ref{eqeff})), the representation of the $\sigma_{\rm ext}$ operator, and the truncation of many-body effects in the effective Hamiltonian. These challenges are typically addressed through: (i) a finite commutator expansion based on the Campbell–Hausdorff theorem; (ii) an approximation of the $\sigma_{\rm ext}$ operator using the unitary coupled cluster formalism, which leverages cluster amplitudes obtained from the standard CCSD simulations; and (iii) the restriction of the effective Hamiltonian to one- and two-body terms.\\

\subsection{CC downfolding codes}

The development of coupled-cluster downfolding computational infrastructure has been the focus of intensive efforts to establish an HPC framework capable of leveraging massively parallel GPU-based architectures for tackling real-world chemistry problems. Initial pilot implementations—hand-coded and serial—were instrumental in defining the hierarchical structure of the downfolded Hamiltonians based on double unitary coupled-cluster (DUCC) approximations, including classes of  DUCC(2) and DUCC(3) approximate models \cite{doublec2022} (see also Appendix \ref{appendix:a7}). These early implementations assumed that all occupied orbitals were active \cite{10.1021/acs.jctc.0c00421,Bauman2021variational}. 

\begin{figure}
    \centering  \includegraphics[width=0.99\linewidth]{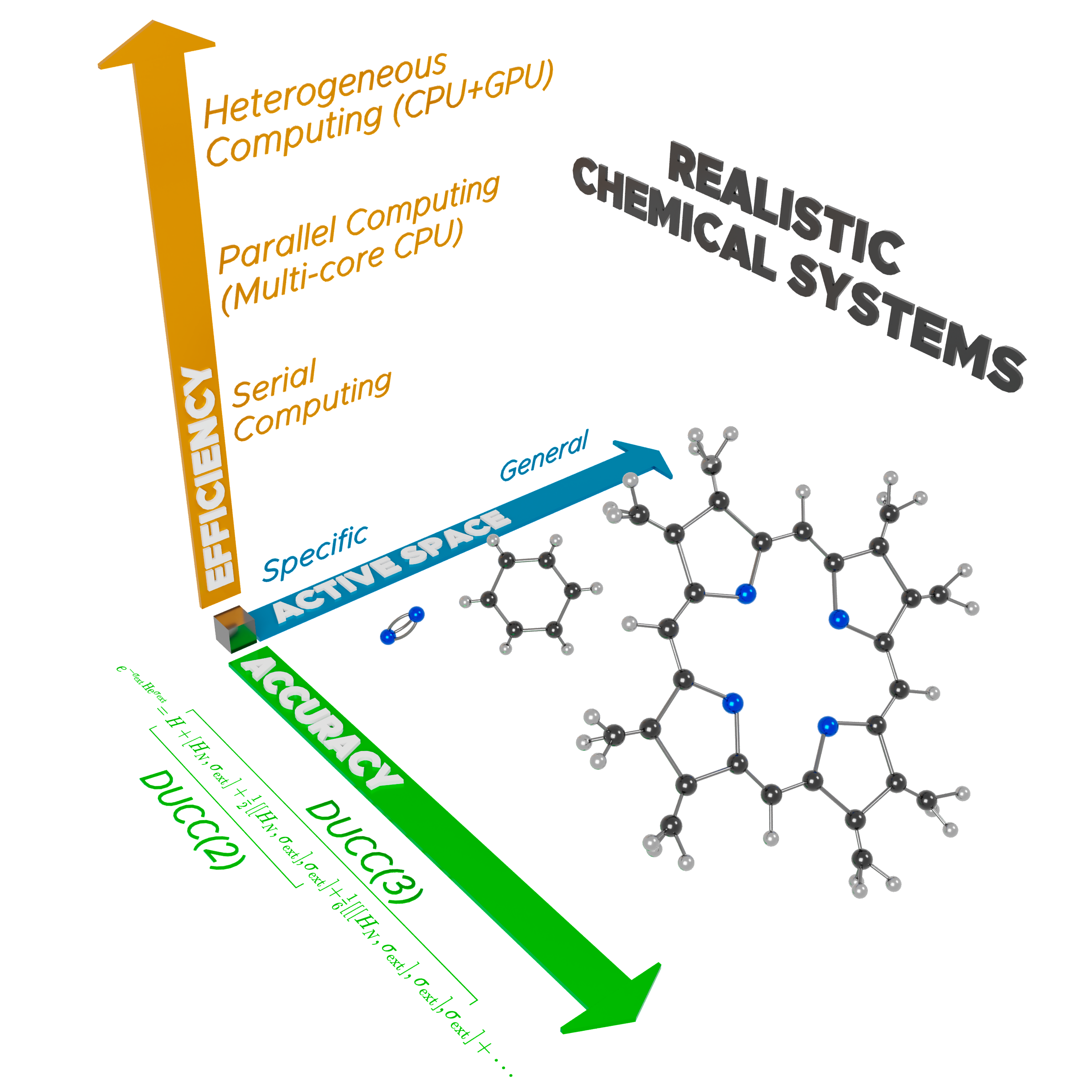}
    \caption{Overview of the evolution of the DUCC family of approximations and their classical implementations for downfolded Hamiltonians. Current implementations leverage heterogeneous HPC architectures, enabling routine application of advanced DUCC(3) approximations to large systems with arbitrarily sized active spaces.}
    \label{fig:ducc_family}
\end{figure}

To extend these models to larger systems and enable their use in simulations on quantum hardware, two major challenges were addressed:
(1) the development of parallel CC downfolding implementations with efficient utilization of GPU technology, and
(2) the formulation of CC downfolded Hamiltonians that support arbitrary active space sizes, thereby enabling simulations of large systems on existing quantum hardware.
The evolution of the CC downfolding software is schematically illustrated in Fig.~\ref{fig:ducc_family}.

In this paper, we utilize the DUCC(3)-type approximation in which the effective Hamiltonians are 
defined by single, double, and partially triple commutators (see the A7 approach in Ref.~\onlinecite{doublec2022}; see also Appendix \ref{appendix:a7}) with external cluster amplitudes 
sourced from CCSD calculations. To obtain the scalar, one- and two-body elements of the 
effective Hamiltonian, we used the parallel DUCC implementation developed in the ExaChem electronic structure code (available on public GitHub
repositories located at \url{https://github.com/ExaChem/exachem}). The many-body form of the DUCC(3) effective Hamiltonian required evaluating just over 1000 Hugenholtz diagrams. Implementing these numerous terms was facilitated by SymGen (\url{https://github.com/npbauman/SymGen}), a symbolic algebra tool that automatically derives tensor expressions corresponding to the second-quantized form of operators typically encountered in various CC methodologies \cite{bylaska2024electronic}. The output equations of SymGen are translated into the TAMM (Tensor Algebra for Many-Body Methods library \cite{mutlu2023tamm}) format utilized in ExaChem, thereby streamlining the translation of tensor contractions to efficient parallel code and eliminating the need for the hand derivation and implementation of hundreds or thousands of diagrams. In contrast to earlier implementations of CC downfolding, where all occupied orbitals were assumed to be active, the most recent version of the CC downfolding software enables the construction of effective Hamiltonians for arbitrary active spaces.  

\subsection{Database of downfolded Hamiltonians }
The effective Hamiltonians of the molecular systems in this study are hosted on a public library repository (\url{https://github.com/npbauman/DUCC-Hamiltonian-Library}). The downfolded Hamiltonian library contains other molecular systems with varying active space sizes, different DUCC approximations, and different parent basis sets used to source the external CC amplitudes. Each system contains the ExaChem input and outputs and the fermionic effective Hamiltonians in various file formats, including raw text, FCIDUMP, and YAML based on the Broombridge schema\cite{low2019qnwchemtoolsscalable}. Additional information about the effective Hamiltonians can also be found in the library, such as orbital energies and corresponding RHF and FCI energies for easy comparison and validation.

\subsection{Simulation software} 
For classical validations, we employed SV‑Sim\cite{li2021svsim}, a scalable state‑vector simulator explicitly designed for large, deep quantum circuits on CPU and GPU clusters, under the NWQSim repository (\url{https://github.com/pnnl/nwq-sim}). NWQSim~\cite{li2021svsim, li2020density, suh2024simulating, li2024tanq} replaces conventional MPI traffic with a shared memory model known as Partitioned Global Address Space (PGAS). It relies on OpenSHMEM, NVSHMEM, and ROCSHMEM for CPU, NVIDIA GPU, and AMD GPU one-sided communication, enabling fine‑grained, low‑latency exchange of wavefunction data distributed across thousands of CPU/GPU devices. The simulator offers both C++ and Python front‑ends and plugs into mainstream quantum‑software stacks, like Qiskit, Q\#, and XACC so the same code that generates our variational quantum‑chemistry circuits could be executed unchanged on classical hardware. 

For implementing our noiseless classical simulation of the VQE algorithm with the UCCGSD ansatz, we have utilized the PennyLane Python library.

\subsection{Solvers} 

Under the standard UCCSD ansatz \cite{pal1984use, unitary1, hoffmann1988unitary, kutzelnigg1991error, unitary2, sur2008relativistic, cooper2010benchmark, evangelista2019exact, anand2022quantum}, the cluster operator is constructed by summing over all the operators that generate single and double excitations between the occupied and virtual orbitals. A direct generalization of this approach, known as unitary coupled cluster generalized singles and doubles (UCCGSD), involves lifting the restriction that excitations must be between occupied and virtual orbitals. This results in a cluster operator that can have as many parameters as the number of two-particle integrals in the electronic Hamiltonian \cite{Nooijen2000eigenstates, Nakatsuji2000structure, VanVoorhis2001coupled, Piecuch2003exactness, Ronen2003eigenstates, Mukherjee2004comments}. The increased number of parameters in the UCCGSD ansatz has been shown to produce high-accuracy results when applied in VQE algorithms \cite{lee2018generalized, GreeneDiniz2021generalized, Bauman2021variational}.

The implementations of ADAPT-VQE and qubit-ADAPT-VQE we employed in SV-Sim are based on Refs.~\onlinecite{avqe} and \onlinecite{qavqe}. ADAPT-VQE adaptively approximates a Hamiltonian’s ground-state energy by iteratively selecting single and double excitation operators that yield the largest energy improvement from an operator pool. This approach produces shallower circuits compared to VQE with a UCCSD ansatz\cite{avqe}. To further reduce two-qubit gates, qubit-ADAPT-VQE limits its operator pool to Pauli strings derived from trotterized fermionic operators\cite{qavqe}. Both methods significantly cut quantum-resource requirements relative to UCCSD VQE while maintaining comparable accuracy.

Inspired by ADAPT-VQE’s adaptive selection and the generator-coordinate method, ADAPT-GCIM uses the same procedure to iteratively choose UCC excitation generators from the same operator pool as fermionic ADAPT-VQE. Rather than optimizing their parameters, it employs these fixed generators to construct a non-orthogonal, overcomplete many-body basis. Projecting the system Hamiltonian into this basis yields an effective Hamiltonian whose generalized eigenvalue problem produces ground- and excited-state energies\cite{zheng2023quantum,zheng2024unleashed}. Although ADAPT-GCIM requires no parameter optimization, a few classical optimization rounds can greatly accelerate convergence\cite{zheng2024unleashed}. We denote this variant ADAPT-GCIM$(x,y)$, where up to $y$ optimization rounds occur for every $x$ ansatz-selection iterations.

A concise overview of the UCCGSD formalism, VQE variants, the GCM framework, and quantum methods inspired by GCM is provided in Appendix \ref{appendix:algs}.

\subsection{Experiments on quantum computers}
\label{sub:hardware}

To confirm the feasibility of the downfolded Hamiltonians and qubit-ADAPT-VQE implementation on real quantum devices, we performed one-shot evaluations of qubit-ADAPT-VQE circuits on a trapped-ion (Quantinuum H1-1) and two superconducting (IBM $ibm\_marrakesh$, $ibm\_kingston$) devices for benzene (cc-pVTZ) and FBP (cc-pVDZ), as previously listed in Table~\ref{tab:hardware}. Additionally, Appendix~\ref{appendix:hardware-spec} details device specifications. Due to credit limits, we did not run ZNE experiments on H1‑1 for FBP.

Both Hamiltonians were mapped to Pauli strings via Jordan–Wigner transformation, grouped by qubit-wise commutativity, and truncated by retaining the groups with the largest sum of coefficient magnitudes to fit within our quantum credit limits. We chose 39 of 116 groups (143/371 strings) for benzene and 59 of 210 groups (311/735 strings) for FBP. This truncation introduced extra ground-state energy errors of $0.4$ milliHartree for benzene and $5.9$ milliHartree for FBP.

The evaluated qubit-ADAPT-VQE circuits used classically selected Pauli evolution ansatzes and classically optimized parameters over \textit{untruncated} Hamiltonians. For benzene, we ran circuits in the $4^{th}$ iteration (4 Pauli evolution ansatzes) on both the H1‑1 emulator and device, and circuits in the $2^{nd}$ iteration (2 Pauli evolution ansatzes) on $ibm\_marrakesh$. For FBP, we executed circuits in the $3^{rd}$ iteration (3 Pauli evolution ansatzes) on the H1‑1 emulator, H1‑1 device, and $ibm\_kingston$.

\begin{figure}[t]
    \centering    \includegraphics[width=0.99\linewidth]{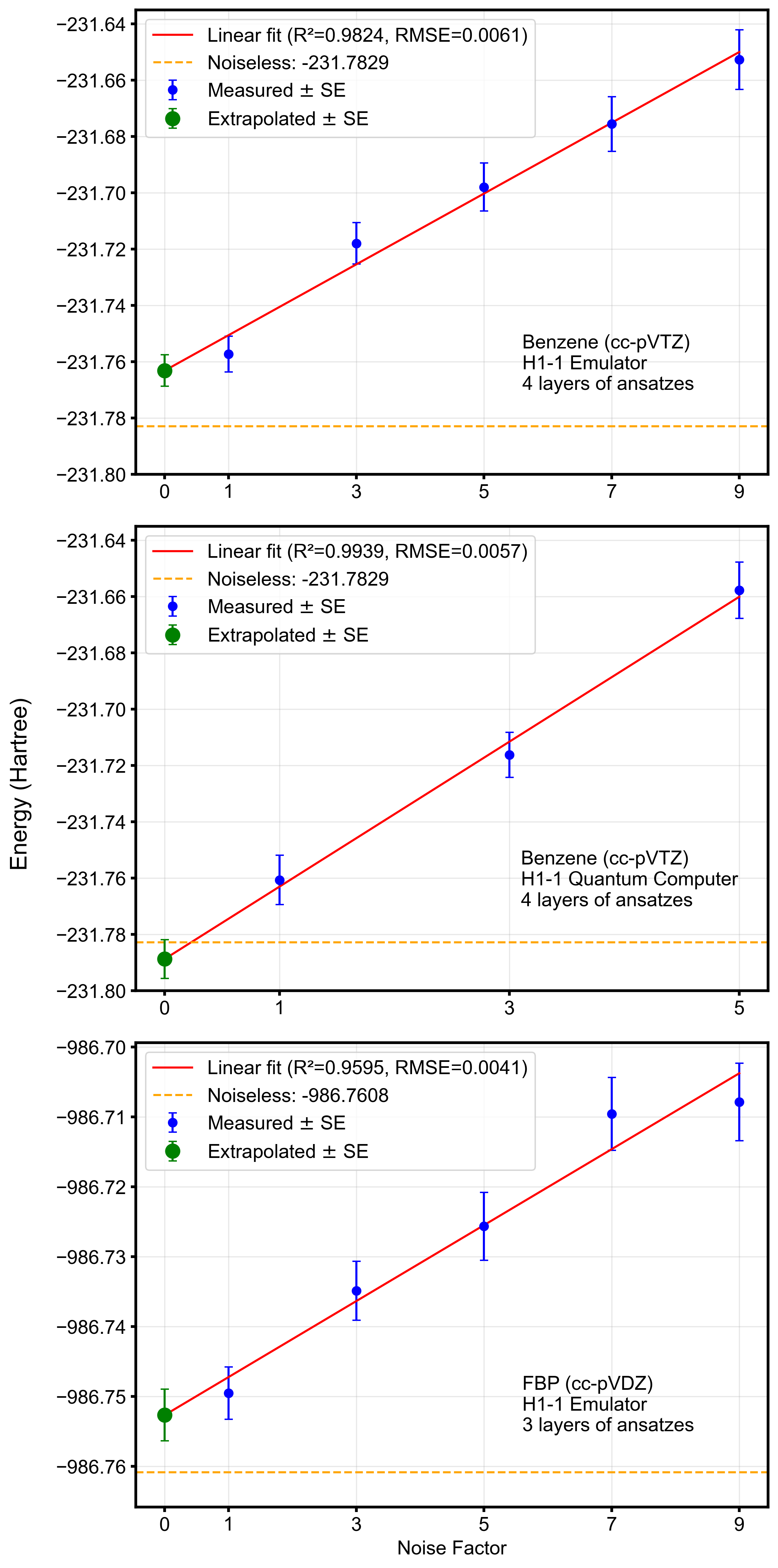}
    \caption{Energy estimates from the Quantinuum H1-1 emulator and quantum hardware for benzene (ccp-VTZ) and FBP, along with extrapolated values. Error bars denote the standard error (SE) for all measured and extrapolated points. Each plot also reports the coefficient of determination ($R^2$) and the root-mean-square error (RMSE).}
    \label{fig:zne_comb}
\end{figure}

In experiments on Quantinuum emulator and hardware, ZNE method, originally developed in Ref.~\citenum{zne_origin}, greatly improved the solution quality. Specifically, our implementation of ZNE is modifyied from the ``MP after'' method in Ref.~\citenum{half_zne} for trapped-ion hardware. We inserted extra 2-qubit gates only in the second half of each circuit. The noise factor equals the ratio of 2-qubit gates in the amplified versus unamplified ansatz. Figure~\ref{fig:zne_comb} shows linear regressions of energy versus noise factor; surprisingly, the extrapolated energy from the H1‑1 device aligns more closely with the noiseless value than that from the emulator. Each circuit ran with 1,024 shots, for a total of 119,808 shots for benzene (including noise-amplified circuits in ZNE) and 60,416 shots for FBP (unamplified only, since ZNE was not performed).

For experiments on IBM quantum computers, we applied Qedma’s QESEM error-mitigation software via Qiskit Function~\cite{qesem}. In a job submission, this software characterizes device noise, transpiles circuits in a noise‑aware manner, and yields unbiased expectation estimates from error-suppressed measurements. Because QESEM outperformed Qiskit Runtime’s built-in ZNE and other error mitigation methods in our trials, we report only QESEM results in Table~\ref{tab:hardware}. Setting the default precisions for QESEM to 0.02, the benzene experiments accumulated 1,927,202 shots, while FBP used 636,200 shots.

\section{Outlook}

Our simulations indicate that recently introduced coupled cluster downfolding techniques are highly effective for obtaining high-accuracy results for realistic chemical systems, even with the limited quantum resources provided by current hardware. The many-body structure of the effective Hamiltonians can effectively capture dynamical correlation effects and extend beyond the limited applicability of active-space models that utilize bare Hamiltonians.

Across all benchmarks considered in this study, we observed that the ground-state energies  of the effective Hamiltonians evaluated using quantum hardware exceeded the accuracy of energies of the approximate CC formulations  (specifically, the CCSD approach used here) employed to provide external amplitudes needed to construct effective Hamiltonians. This effect, which we referred to as the {\it accuracy amplification}, provides additional motivation to align downfolding methods with the continuous advancement of classical and quantum hardware. As quantum computers become capable of handling larger active spaces, further increases in accuracy can be expected.

In this work, we utilized a simple form of the effective Hamiltonian  based on rank-1 and rank-2 (and the simplest terms contributing to rank-3) commutator expansions, using external singly and doubly excited cluster amplitudes stemming from the conventional CCSD simulations. We restricted the many-body components to one- and two-body interactions. This simplified approach leaves room for future improvements in energy accuracy. For example, in the recent study \cite{singh2025qubit}, we demonstrated that the inclusion of three-body interactions in effective Hamiltonians leads to significant improvements in ground-state properties, especially in strongly correlated regimes. \
Additionally, problems defined by downfolded Hamiltonians can be effectively integrated with other quantum solvers recently discussed in the literature
\cite{weaving2025contextual,zhao2025quantum}.


We believe that the adoption of coupled cluster downfolding techniques can accelerate the use of quantum computers in delivering accurate ground-state energies - crucial for understanding a broad range of chemical processes - and enable more efficient utilization of existing hardware in terms of the accuracy of calculated energies.



\section{Author Contributions}

N.P.B. and A.P. developed scalable code for generating downfolded Hamiltonians. M.Z. and C.L. performed hardware simulations and conducted error mitigation analysis. N.M.M. and B.P. contributed to the development of the solvers. A.L. and K.K. conceptualized the research ideas and overall design. All authors analyzed the data and contributed to writing the manuscript.



\section{Acknowledgements}
M.~Z., N.~M.~M., and K.~K. were supported by the ``Embedding QC into Many-body Frameworks for Strongly Correlated  Molecular and Materials Systems'' project, which is funded by the U.~S. Department of Energy, Office of Science, Office of Basic Energy Sciences (BES), the Division of Chemical Sciences, Geosciences, and Biosciences under FWP 72689. A.~L. and K.~K. acknowledge the support from 
Quantum Science Center (QSC), a National Quantum Information Science Research Center of the U.S. Department of Energy (under FWP  76213).
B.~P. acknowledges the support from the Early Career Research Program by the U.S. Department of Energy, Office of Science, under Grant No. FWP 83466.
NPB acknowledges the support from the
“Transferring exascale computational chemistry to cloud computing environment and emerging hardware technologies (TEC4)” project, which is funded by the U.S. Department of Energy, Office of Science, Office of Basic Energy Sciences, the Division of Chemical Sciences, Geosciences, and Biosciences (under Grant No. FWP 82037).
This research was also supported by the Quantum Algorithms and Architecture for Domain Science Initiative (QuAADS), under the Laboratory Directed Research and Development (LDRD) Program at Pacific Northwest National Laboratory (PNNL). 
This research used resources of the Oak Ridge Leadership Computing Facility, which is a DOE Office of Science User Facility supported under Contract DE-AC05-00OR22725. This research used resources of the National Energy Research Scientific Computing Center (NERSC), a U.S. Department of Energy Office of Science User Facility located at Lawrence Berkeley National Laboratory, operated under Contract No. DE-AC02-05CH11231. The Pacific Northwest National Laboratory is operated by Battelle for the U.S. Department of Energy under Contract DE-AC05-76RL01830.

\section{Competing interests}
All authors declare no financial or non-financial competing interests. 

\section{Data and code availability}
The scalar, one- and two-body elements of the 
effective Hamiltonians were obtained using the parallel DUCC implementation developed in the ExaChem electronic structure code (\url{https://github.com/ExaChem/exachem}). The many-body form of the DUCC(3) effective Hamiltonian was derived and implemented in ExaChem using the symbolic algebra tool SymGen (\url{https://github.com/npbauman/SymGen}). The effective Hamiltonians of the molecular systems, as well as other data and the code for related numerical simulations, are hosted on a public repository (\url{https://github.com/npbauman/DUCC-Hamiltonian-Library}). 


\appendix

\section{Approximate forms of the downfolded Hamiltonians} 
\label{appendix:a7}

Eq. \eqref{eqeff} can be rewritten as 
\begin{equation}
        H^{\rm eff} = (P+Q_{\rm int}) \bar{H}_{\rm ext} (P+Q_{\rm int})
\label{equivducc}
\end{equation}
where the external Hamiltonian $\bar{H}_{\rm ext}$ is defined as
\begin{equation}
        \bar{H}_{\rm ext} =e^{-\sigma_{\rm ext}}H e^{\sigma_{\rm ext}}.
\label{duccexth}
\end{equation}
Using the Baker--Campbell--Hausdorff (BCH) formula, the $\bar{H}_{\rm ext}$ can be further expanded in the form of a non-terminating expansion:
\begin{equation}
\bar{H}_{\rm ext} = H + [H,\sigma_{\rm ext}] + \frac{1}{2}[[H,\sigma_{\rm ext}],\sigma_{\rm ext}] + \ldots 
\label{bch}
\end{equation}
The DUCC(3)-A(7) approximation is defined by finite rank commutator expansion based on the BCH formula above, i.e., 
\begin{widetext}
\begin{equation}
    \bar{H}_{\rm ext}^{\rm A(7)} = H+[H_N,\sigma_{\rm ext}]+\frac{1}{2}[[H_N,\sigma_{\rm ext}],\sigma_{\rm ext}]+\frac{1}{6} [[[F_N,\sigma_{\rm ext}],\sigma_{\rm ext}],\sigma_{\rm ext}]
    \label{a7exp}
\end{equation}
\end{widetext}
where $H_N$ and $F_N$ are the normal product forms of the Hamiltonian and the Fock operator.
The many-body form of $\bar{H}_{\rm ext}^{\rm A(7)}$ is limited to scalar and one- and two-body interactions.

\section{CC downfolding: many-body structure of the effective Hamiltonian} 
\label{appendix:heff}


The CC downfolding theory describes a quantum system composed of two or more interacting components. In the simplest case, the quantum system $S$
consists of two interacting parts, $X$ and $Y$, represented by active and inactive spin-orbitals, respectively. CC downfolding can be used to construct an effective Hamiltonian for subsystem $X$ that reproduces the energy of the entire system  $S$.

The CC  downfolding  addresses several issues commonly associated with traditional downfolding/embedding procedures. 
Firstly, because it is firmly grounded in coupled cluster theory, which features a well-defined hierarchy of approximations, it avoids the so-called ``double-counting'' problem in the treatment of electron correlation effects. This issue frequently arises in approaches where correlation effects cannot be rigorously categorized. It is particularly common in embedding methods that employ various parameterizations (e.g., wave function, density matrix, or electron density) of quantum systems, where translating correlation effects between different approximations is often not feasible,  especially when combining methodologies in which establishing a consistent hierarchy of correlation effects is inherently impossible.

\begin{figure}
    \includegraphics[width=\linewidth]{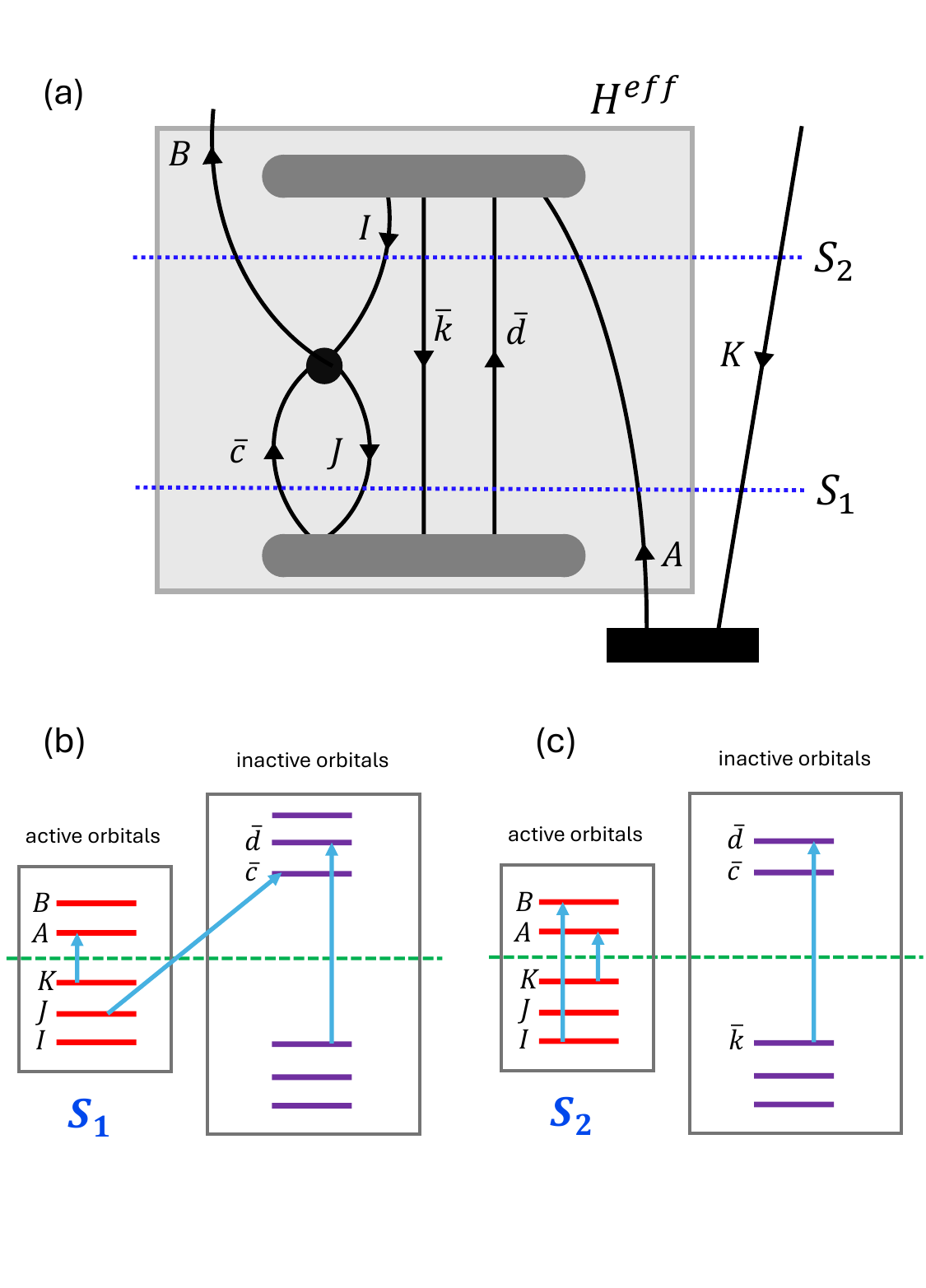}
    \caption{
(a) Typical Hugenholtz diagram contributing to the one-body part of the effective Hamiltonian. Gray ovals represent amplitudes defining the $\sigma_{\rm ext}$/$\sigma_{\rm ext}^\dagger$ operators. The black dot represents antisymmetrized vertices corresponding to pairwise interactions in the bare Hamiltonian. The letter indices $I, J, K$ / $A, B$ ($\bar{k}$ / $\bar{c}, \bar{d}$) denote occupied/virtual active (inactive) spin-orbitals. This diagram contributes to the $a_B^\dagger a_A$ excitation, which commutes with the active-electrons number operator, $n_{\rm act} = \sum_P a_P^\dagger a_P$, where $P$ is a generic active spin-orbital index. The intermediate states $S_1$ and $S_2$, however, involve configurations in the Hilbert space that correspond to the transfer of electrons between active (embedded system) and inactive (environment) spin-orbitals needed to correlate interacting subsystems $X$ and $Y$.
The diagram is represented in action of singly excited determinant $|\Phi_K^A\rangle = a_A^{\dagger} a_K |\Phi\rangle$.
(b) a symbolic representation of triple excitation, with respect to the reference function $|\Phi\rangle$,  defining the intermediate $S_1$ state. One can notice that the active electron in occupied active spin-orbital $J$ is promoted to virtual inactive spin-orbital $\bar{c}$, (c) a symbolic representation of triple excitation, with respect to the reference function $|\Phi\rangle$,  defining the intermediate $S_2$ state.
}
    \label{fig:diagram}
\end{figure}

Secondly, although effective Hamiltonians conserve the number of active electrons during simulations, i.e., 
\begin{equation}
   [H^{\rm eff}, n_{\rm act}] = 0 \;,
   \label{cummno}
\end{equation}
where $n_{\rm act}$ is active-electrons number operator, 
the diagrammatic analysis (as shown in Fig.~\ref{fig:diagram}) reveals that CC downfolding facilitates explicit correlation between the active space and its orthogonal complement. The diagram in Fig.~\ref{fig:diagram} illustrates a scenario where the number of active-space electrons is preserved in the initial and final states; however, in the intermediate states, electrons can hop between the active and inactive spin-orbitals. 

\section{UCCGSD ansatz and VQE and GCM algorithms} 
\label{appendix:algs}

In the standard UCCSD ansatz, the wave function is given by, 
\begin{equation}
    \ket{\Psi} = e^{T-T^{\dagger}}\ket{\phi} \label{eq:uccsd}
\end{equation}
where the cluster operator, $T$, is truncated at the level of singles and doubles excitations,
\begin{equation}
    T = \sum_{i, a} t_i^a a_a^{\dagger} a_i + \sum_{i<j,a<b} t_{ij}^{ab} a_a^{\dagger} a_b^{\dagger} a_j a_i. \label{eq:sdexc}
\end{equation}
Here, the indices $i, j$ run over occupied spin orbitals and the indices $a, b$ run over the virtual orbitals. The cluster operator for the UCCGSD ansatz is constructed in an analogous manner,
\begin{equation}
    \Gamma = \sum_{p, q} \gamma_p^q a_p^{\dagger} a_q + \sum_{p<q,r<s} \gamma_{pq}^{rs} a_p^{\dagger} a_q^{\dagger} a_s a_r,
\end{equation}
with the key distinction that in this case the indices $p, q, r$ and $s$ run over all spin orbitals.

In ADAPT-VQE, individual single and double excitation operators are put in a pool of operators. Denote a single and a double excitation operator as
\begin{align}
    A_{p_k,q_k} &= a_{p_k}^\dagger a_{q_k} - a_{q_k}^\dagger  a_{p_k}, \label{eq:apq}\\
    A_{p_k,r_k,q_k,s_k} &= a_{p_k}^\dagger a_{r_k}^\dagger a_{q_k} a_{s_k} - a_{s_k}^\dagger a_{q_k}^\dagger a_{r_k} a_{p_k}.\label{eq:aprqs}
\end{align}
where $k$ indexes the operator. Under Jordan--Wigner mapping, their evolutions in qubit space are
\begin{align}
    &\exp(\theta A_{p_k,q_k}) = \exp\bigg( - {\rm i} \frac{\theta}{2}
    (X_{q_k}Y_{p_k} - Y_{p_k}X_{q_k}) \prod_{l=q_k+1}^{p_k-1} Z_l
    \bigg), \label{eq:singlepauli} \\
    &\exp(\theta A_{p_k,r_k,q_k,s_k}) = \exp\bigg( - {\rm i} \frac{\theta}{8}
    (X_{q_k}Y_{s_k}X_{p_k}X_{r_k} + Y_{q_k}X_{s_k}X_{p_k}X_{r_k} \notag \\
    &~~+Y_{q_k}Y_{s_k}Y_{p_k}X_{r_k} + Y_{q_k}Y_{s_k}X_{p_k}Y_{r_k} - X_{q_k}X_{s_k}Y_{p_k}X_{r_k} - X_{q_k}X_{s_k}X_{p_k}Y_{r_k} \notag \\
    &~~ -Y_{q_k}X_{s_k}Y_{p_k}Y_{r_k} - X_{q_k}Y_{s_k}Y_{p_k}Y_{r_k}) \prod_{k=p_k+1}^{r_k-1} Z_k \prod_{l=q_k+1}^{s_k-1} Z_l
    \bigg),  \label{eq:doublepauli}
\end{align}
and can be constructed in quantum circuits\cite{barkoutsos2018quantum, PhysRevA.102.062612}. Without the loss of generality, we use $e^{\theta_k A_k}$ to denote $k^{th}$ ansatz in the operator pool, regardless from single or double excitation operator. Let the $l^{th}$ iteration of ADAPT-VQE prepares the state
\begin{align}
    \ket{\psi^{(l)}} = \prod_{k = 1}^{l}  e^{\theta_k A_k} \ket{\psi_{HF}} \label{eq:avqestate}
\end{align}
where $\ket{\psi^{(0)}}:=\ket{\psi_{HF}}$ is the Hartree--Fock state. At $(l+1)^{th}$ iteration, the gradient $\bra{\psi^{(l)}} \left[H,A_i\right] \ket{\psi^{(l)}}$ are evaluated for all $k$ from 1 to $N$ to selected the $A_k$ that gives the largest magnitude of the gradient, where $H$ is the Hamiltonian of interest after Jordan-Wigner mapping and $N$ is the number of operators in the pool. Then, the new state
\begin{align}
    \ket{\psi^{(l+1)}} =  e^{\theta_{l+1} A_{l+1}} \prod_{k = 1}^{l}  e^{\theta_k A_k} \ket{\psi_{HF}}
\end{align}
is minimized over VQE objective function $\bra{\psi^{(l+1)}}H \ket{\psi^{(l+1)}}$ for optimal parameter values and checked for convergence. The convergence criteria can be set to the norm of the gradient.

In this work, we adopted the qubit-ADAPT-VQE algorithm\cite{tang2021qubit} to further simplify the ansatz of the variational algorithm.
Specifically, instead of using the linear combination of Pauli strings mapped from fermionic excitation operators, the operator pool in qubit-ADAPT-VQE only selects the individual Pauli strings from the summations on exponents in Eqs.~\eqref{eq:singlepauli} and~\eqref{eq:doublepauli} with an odd number of Pauli $Y$.
The long $Z$ tails of the Pauli strings are also removed for hardware efficiency.
This greatly reduces the circuit depth of each Pauli rotation and the number of 2-qubit gates. In addition, Ref.~\citenum{tang2021qubit} proves that using a minimum operator pool leads to a number of ansatz that grows linearly with the number of qubits, which reduces the total number of measurements for each ADAPT iteration.

Another wavefunction approximation technique is the Generator Coordinate Method (GCM). Rooted in the variational principle, GCM provides a powerful framework for describing collective excitations and quantum fluctuations within a many-body system. At its core, GCM constructs the many-body wave function as a superposition of a continuous set of ``generator states,'' each characterized by a specific value of one or more collective coordinates. These generator states are typically built from mean-field solutions (e.g., Hartree--Fock or Hartree--Fock--Bogoliubov) constrained to specific values of these collective coordinates. The coefficients of this superposition are then determined by solving the Hill-Wheeler equation, which effectively projects the many-body problem onto a lower-dimensional collective space, thereby capturing correlations beyond the static mean-field picture.

While highly successful in its traditional applications, the full GCM formulation can become computationally demanding, especially when dealing with a large number of generator coordinates or complex basis sets. This computational bottleneck has motivated the development of more flexible and computationally efficient alternatives. One such development is the Generator-Coordinate-Inspired Method. GCIM retains the spirit of GCM by constructing a variational wave function as a linear combination of non-orthogonal states, but it offers greater flexibility in the choice and construction of these ``basis states'' (or configurations) compared to the strictly collective-coordinate-dependent generator states of traditional GCM. This allows for a more general exploration of the many-body Hilbert space, potentially incorporating configurations that are not easily mapped to simple collective coordinates, while still leveraging the variational principle.

To further enhance the efficiency and adaptability of GCIM, particularly in scenarios where the most relevant configurations are not a priori obvious, we have recently introduced an adaptive form of GCIM employing a customized operator pool. This adaptive GCIM methodology dynamically constructs the variational basis by iteratively adding configurations that are deemed most important for accurately capturing the system's ground state or low-lying excited states. This is achieved by systematically applying a set of carefully chosen quantum operators from a ``customized operator pool'' to an initial reference state or an existing set of configurations. The selection of these operators, and thus the generated configurations, is guided by a criterion that prioritizes those contributing most significantly to the correlation energy or that are most strongly coupled to the current variational space. This adaptive strategy allows for a highly efficient exploration of the relevant Hilbert space, focusing computational resources on the most impactful correlations and offering a significant advantage over methods that rely on pre-defined or exhaustively generated basis sets, thereby extending the applicability of generator-coordinate-inspired approaches to a broader range of complex quantum many-body problems.

As in Ref.~\citenum{zheng2024unleashed}, ADAPT-GCIM can employ the same operator pool, which consists of excitation operators in Eq.~\eqref{eq:apq} and Eq.~\eqref{eq:aprqs}, and the state preparation is the same as in Eq.~\eqref{eq:avqestate}. Rather than optimizing parameters for a VQE objective function, it fixes all parameters a pre-defined constant, then iteratively expands a non-orthogonal, overcomplete basis set using prepared states $\{\ket{\psi_{HF}}, \ket{\psi^{(1)}}, \dots\}$. Projecting the Hamiltonian $H$ onto this basis gives an effective Hamiltonian $\mathbf{H}$ and overlap matrix $\mathbf{S}$, leading to the generalized eigenvalue problem $\mathbf{H} f = \epsilon \mathbf{S} f$, whose smallest eigenvalue can be an accurate estimation of the ground-state energy of $H$, while some of the other eigenvalues correspond to the excitation-state energies.

\section{Hardware Specifications}
\label{appendix:hardware-spec}

Table~\ref{tab:spec} lists the specifications of the quantum hardware we used, as reported by the vendors. Quantinuum’s data are averages obtained via benchmarking tests in Ref.~\citenum{qtmh11}. IBM’s data comes from its daily calibration, and the values in the table are averages that exclude failed gates (whose error rates equal to 1)~\cite{ibmmachines}.

\begin{table}
    \centering
    \caption{Quantinuum and IBM hardware specifications}
    \label{tab:spec}
    \begin{tabular}{lccc} \hline \hline \\
                             &  Quantinuum & \multicolumn{2}{c}{IBM}   \\
      Category               & H1-1           & $kingston$    & $marrakesh$ \\ \hline \\[-0.1cm]
      1q gate error          & 1.80e-5        & 3.61e-4          & 5.86e-4 \\ [0.2cm]
      2q gate error          & 9.73e-4        & 5.45e-3 (CZ)     & 6.09e-3 (CZ) \\ [0.2cm]
      Pr(meas. 0 prep. 1)    & 3.43e-3        & 2.44e-2          & 4.10e-2  \\ [0.2cm]
      Pr(meas. 1 prep. 0)    & 1.22e-3        & 1.60e-2          & 1.82e-2  \\ [0.2cm]
      T1                     & Not Available  &259.08 $\mu s$    & 202.20 $\mu s$ \\ [0.2cm]
      T2                     & Not Available  &167.09 $\mu s$    & 130.90  $\mu s$ \\ [0.2cm]
      Data Date              & May 2, 2025    & \multicolumn{2}{c}{June 16, 2025} \\ 
      \hline \hline
    \end{tabular}
\end{table}


\bibliography{references}

\end{document}